\documentclass[seceq,jjapprn]{jjap}
\usepackage[dvips]{epsfig}

\title{Direct X-ray Imaging of $\mu$m precision using Back-Illuminated CCD}

\author{Emi {\sc Miyata}$^{1}$, Masami {\sc Miki}$^{1}$, Hiroshi {\sc
Tsunemi}$^{1}$, Junko {\sc Hiraga}$^{1}$, Hirohiko {\sc Kouno}$^{1}$,
and Kazuhisa {\sc Miyaguchi}$^{2}$ }

\inst{
$^1$ Department of Earth \& Space Science, 
Graduate School of Science, Osaka University, \\
1-1 Machikaneyama, Toyonaka, Osaka 560-0043, Japan\\
$^2$ Solid State Division, Hamamatsu Photonics K.K., \\
1126-1 Ichino-cho, Hamamatsu City 435-8558, Japan
}

\abst{ A charge-coupled device (CCD) is a standard imager in optical
region in which the image quality is limited by its pixel size.  CCDs
also function in X-ray region but with substantial differences in
performance.  An optical photon generates only one electron while an
X-ray photon generates many electrons at a time.  We developed a method
to precisely determine the X-ray point of interaction with subpixel
resolution.  In particular, we found that a back-illuminated CCD
efficiently functions as a fine imager.  We present here the validity of
our method through an actual imaging experiment. }

\kword{charge-coupled device, X-ray imaging, subpixel spatial
resolution, mesh experiment}

\begin{document} 
 
 \maketitle 
 \sloppy

 Charge-coupled devices (CCDs) are widely used for optical and X-ray
 imaging.  When an X-ray photon is absorbed by photoelectric interaction
 in the photo-sensitive region (depletion region) of the CCD, it
 generates a primary charge cloud consisting of many electrons whose
 number is proportional to the incident X-ray energy $E$.  The size of
 the primary charge cloud is 73\,($E$/2.3
 keV)$^{1.75}$nm~\cite{cloud_size}.  The primary charge cloud expands
 through diffusion process until they reach the potential well of the
 CCD pixel.  Therefore, the final charge cloud after the diffusion
 process, which is much bigger than the primary charge cloud, is
 accumulated into several adjacent pixels forming various types of event
 pattern (``grade'') depending on how they split.  When the entire
 charge is collected into one pixel (no surrounding pixels having any
 signal), it is called a ``single pixel event'' while when it splits
 into more than one pixel, it is called a ``split pixel event''.

 There are two types of CCDs: front-illuminated (FI) CCD and
 back-illuminated (BI) CCD with basically the same structure.  There are
 electrodes on the front side of the CCD.  The depletion region is
 generated below electrodes while the potential well is generated very
 close to electrodes.  Therefore, the charge is collected near the front
 side.  The FI CCD is widely used so far in which photons enter into the
 depletion region from the front side.  The depletion region of the FI
 CCD is covered by electrodes that reduce the detection efficiency for
 low energy X-rays and blue optical light.  On the contrary, in the BI
 CCD, photons enter into the depletion region from the backside of the
 CCD.  The primary charge generated near the backside of the CCD drifts
 toward the potential well near the front side.  A BI CCD is expected to
 have higher detection efficiency than a FI CCD since photons enter into
 the depletion region without passing through the electrodes.

 Tsunemi et al.~\cite{tsunemi97} developed a new technique, ``mesh
 experiment'', that enables us to restrict the X-ray point of
 interaction with a subpixel resolution.  With this technique, Hiraga et
 al.~\cite{hiraga98} directly measured the final charge cloud shape
 accumulated in the potential well through diffusion.  They found that a
 final charge cloud shape could be well represented by a Gaussian
 function.  They also obtained standard deviation, $\sigma$, of the
 final charge cloud to be $0.7 \sim 1.5 ~\mu$m for $1.5\sim 4.5$~keV by
 using a FI CCD.  Based on this experiment, they confirmed that there
 are three parameters tightly coupled together~\cite{tsunemi00}: the
 X-ray point of interaction inside the pixel, the way the final charge
 splits among pixels and the final charge cloud shape.  Any two
 parameters can determine the third one.  The event grade is quite
 easily measured while the final charge cloud shape is difficult to
 obtain.  Currently, they can separately measure it only by using the
 mesh experiment.  Therefore, they can determine the X-ray point of
 interaction with much better resolution than the pixel size for split
 pixel events.  They obtained a position resolution of 0.7~$\mu$m using
 a CCD with pixel size of 12~$\mu$m~\cite{hiraga01}.

 Most of the X-ray photons photo-absorbed inside a FI CCD produce final
 charge clouds of relatively smaller size.  This is due to the fact that
 the photo-absorption occurs near the potential well and a split pixel
 event occurs only when the point of interaction is near the pixel
 boundary.  On the contrary, X-ray interaction in a BI CCD produces
 relatively large size of final charge cloud: $3 \sim 6~\mu$m and most
 of the X-ray photons form split pixel events~\cite{bi_mesh}.  In a BI
 CCD, the primary charge cloud generated near the backside travels
 relatively longer distance before reaching the potential well,
 resulting the diffusion into a large size of the final charge cloud.
 Since the size of the final charge cloud is larger in a BI CCD and the
 shape can be easily obtained for most of the photons, it allows us to
 determine the point of interaction with much greater accuracy.  We
 performed a demonstration that a BI CCD works as a very fine imager.

 The BI CCD employed in this experiment is ``S7030'', manufactured by
 Hamamatsu Photonics Inc.  It consists of $512 \times 122$ active pixels
 of 24~$\mu$m square.  The chip is cooled down to $-100 ^\circ$C so that
 we can run the CCD in a photon counting mode with low noise level.  A
 readout noise level of 5--10 electrons is achieved by using the {\sl
 E-NA} system~\cite{e-na}.  To use the Mo-L X-rays (2.3\,keV), we
 employed a RIGAKU X-ray generator Ultra X-18 with a Mo target, with
 which we found that more than 90~\% of X-ray events form split pixel
 events.  Since the CCD is a full-frame transfer type, we used a
 mechanical shutter in front of the CCD so that X-rays do not enter
 during the readout time (2\,s/frame).  The X-ray intensity is also
 controlled such that there is no pile-up during the exposure time.
 Since it is relatively difficult to prepare very fine X-ray image on
 the CCD, we simply placed two straight metal plates forming a V-shape
 shadow about 2\,mm above the CCD.  In this way, we collected data for
 $\sim$6~ks during which $\sim$ 2 million of photons were accumulated.
 Figure~1 shows an optical image of the V-shape structure as well as the
 X-ray shadow.  The ``V'' has a sharp edge with an opening angle of
 $24.5^\circ$.  The X-ray image in Figure~1 (b) is a sum of all the
 frames obtained.  This is equivalent to a CCD image with a long
 exposure time in a photographic mode.  Each pixel in the X-ray bright
 region contains $\sim$100 of photons.

 We run the CCD in a photon counting mode with which we confirmed that
 most of the X-ray photons formed split events.  Among 2\,million X-ray
 events, we selected split events such that we could improve the X-ray
 point of interaction.  Therefore, we excluded single pixel events and
 two-pixel split events that constitute less than 10~\% of the total
 events.  We can measure the incident X-ray energy by adding signals
 from all the pixels that form the event.  In this way, we selected only
 the Mo-L X-rays so that we could safely assume that the final charge
 cloud shape is a Gaussian function of $\sigma = 6\,\mu$m.  The final
 charge cloud shape is big enough that some charge spills over the pixel
 forming a split event.  The charge distribution among the pixels depend
 on the X-ray point of interaction: the closer to the pixel boundary,
 the more charge splits into the adjacent pixel.  Then, we calculated
 the X-ray point of interaction for each event referring to the charge
 distribution among the event.

 Figure~2 (a) shows an enlarged view of the region near the sharp edge
 in Fig~1 (b).  Each small square corresponds to a CCD pixel of $24~
 \mu$m square.  We see that the sharp edge is rounded by the pixel size.
 Figure~2 (b) is the corresponding image showing the position of
 interaction of each individual photon determined using the final charge
 cloud shape and the way the charge splits among pixels.  We can clearly
 see a sharp structure of the iamge.  Judging from the X-ray wave length
 ($\sim 0.5$\,nm) and the distance between the V-shape structure and the
 CCD, the shadow is blurred by 1\,$\mu$m due to the diffraction.  Taking
 into account this effect, we believe that the position resolution is
 improved to less than $2~\mu$m.  However, the photon density obtained
 is about 100\,photons/pixel that corresponds to
 0.17\,photons/$\mu$m$^2$.  This is too sparse density to practically
 obtain a clear image with $\mu$m scale.  This is an intrinsic problem
 in our method.  The X-ray intensity should be controlled so that
 pile-up does not become serious in $24~\mu$m size pixel while we need
 enough number of photons in $1~\mu$m square region, which is our goal.
 The only method to solve this problem is to speed up readout.  However,
 the readout speed is limited to about 1\,MHz so that we can run the
 system in a very low noise level in order to keep good energy
 resolution.  At present, it is inevitable to spend long time to obtain
 a good spatial resolution.  It may also be a problem to firmly fix the
 target with $1~\mu$m scale during a relatively long exposure time.

 \acknowledgement

 We acknowledge Kazutoshi Yasui for his support of the experiment.
 J.H. is partially supported by JSPS Research Fellowship for Young
 Scientists, Japan.  The manuscript is read by Dr. B. Paul.  This work
 is partly supported by the Grant-in-Aid for Scientific Research by the
 Ministry of Education, Culture, Sports, Science and Technology of Japan
 (13874032, 13440062).

\begin{halffigure}
 \caption{(a) Optical image (12\,mm square) of a V-shape structure on
 the CCD.  (b) Enlarged image near the V-shape structure in X-ray.  A
 sharp edge is seen.}  \label{shadow}
\end{halffigure}

\begin{halffigure}
 \caption{(a) Simple expansion ($300 \times 500 \,\mu$m$^2$) of the
 X-ray image in figure 1.  Squares are CCD pixels of size $24\,\mu$m.
 (b) Same image with improved resolution.  Each dot represents an
 individual X-ray photon.}  \label{image}
\end{halffigure}

\makefigurecaptions

 \clearpage
 
 \includegraphics[clip,scale=.8]{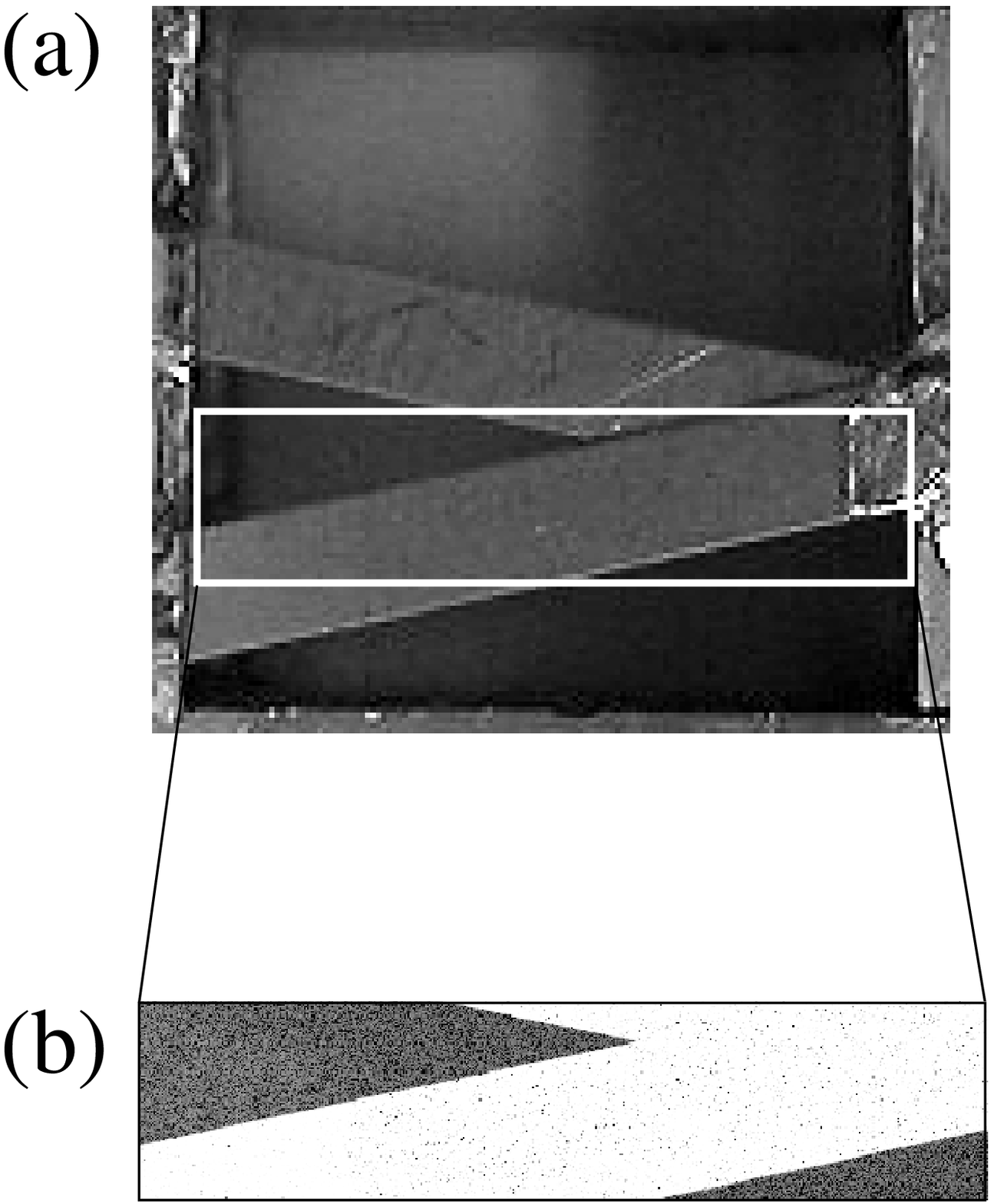}
 
 \clearpage
 
 \includegraphics[clip,scale=.6]{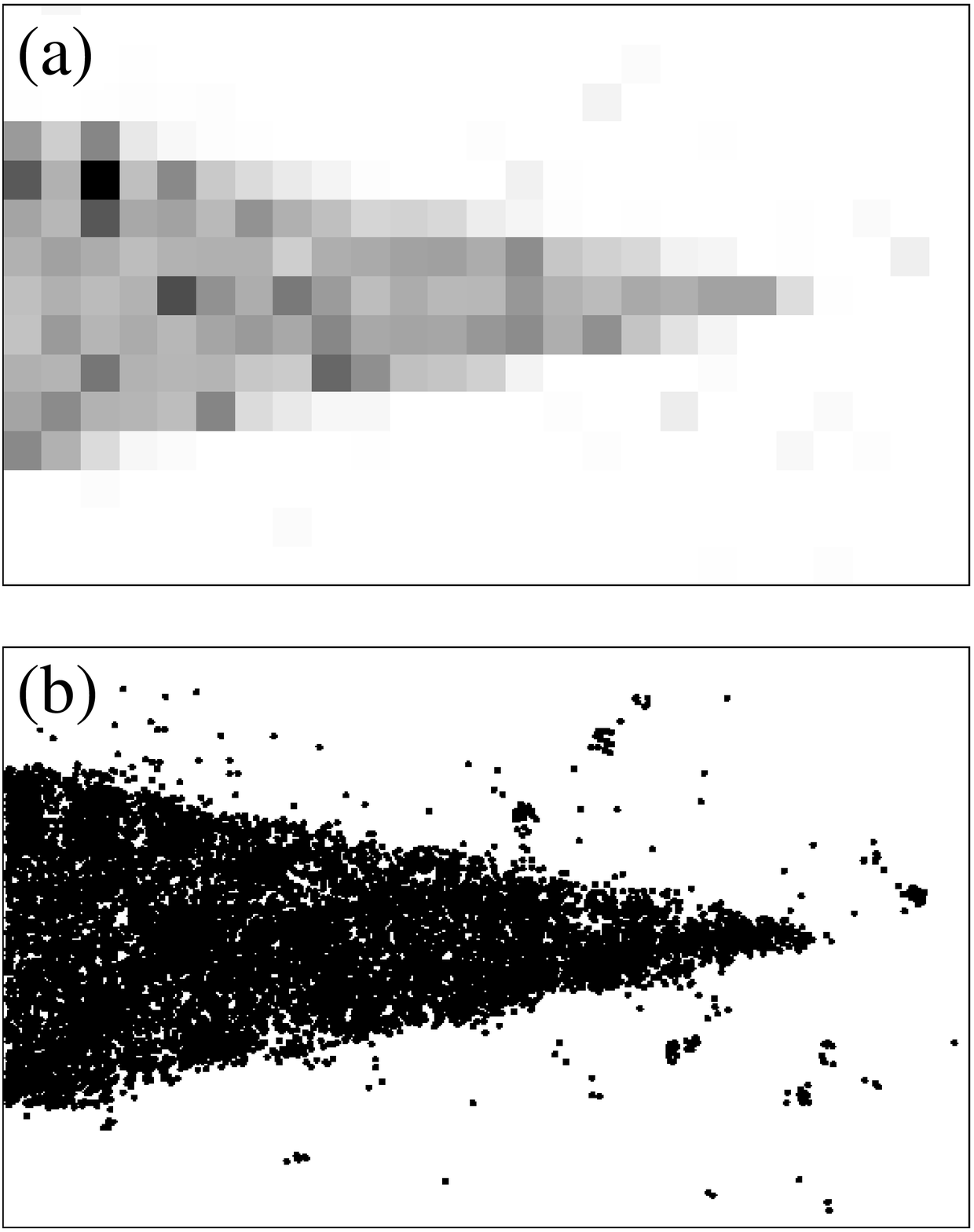}


\begin{thebibliography}{99}

 \bibitem{cloud_size} T.E. Everhart, P.H. Hoff: J. Appl. Phys. {\bf 42}
         (1971) 5837
	 
 \bibitem{tsunemi97} H. Tsunemi, K. Yoshita and S. Kitamoto:
	 Jpn. J. Appl. Phys. {\bf 36} (1997) 2906.

 \bibitem{hiraga98} J. Hiraga, H. Tsunemi, K. Yoshita, E. Miyata and
	 M. Ohtani: Jpn. J. Appl. Phys., {\bf 37} (1998) 4627

 \bibitem{tsunemi00} H. Tsunemi, J. Hiraga and E. Miyata:
	 Nucl. Instrum. \& Methods, (2002) {\bf 477}, 155

 \bibitem{hiraga01} J. Hiraga, H. Tsunemi and E. Miyata:
	 Jpn. J. Appl. Phys., {\bf 40} (2001) 1493

 \bibitem{bi_mesh} E. Miyata et al.: in preparation for publication

 \bibitem{e-na} E. Miyata, C. Natsukari, D. Akutsu, T. Kamazuka, 
	 M. Nomachi, M. Ozaki, 
	 Nucl. Instrum. \& Methods: {\bf 459}  (2001) 157
\end{thebibliography}
 \end{document}